\documentclass[prl,twocolumn,graphicx,amssymb,floatfix]{revtex4}

\usepackage{graphicx}
\begin{document}

\title{The past of a quantum particle. }

\author{L. Vaidman}
\affiliation{ Raymond and Beverly Sackler School of Physics and Astronomy\\
 Tel-Aviv University, Tel-Aviv 69978, Israel}

\begin{abstract}
 Although there is no consensus  regarding the ``reality'' of the past of a quantum  particle, in  situations where there is only one trajectory with nonvanishing quantum wave of the particle between its emission and detection points, it seems ``safe" to associate the past of the particle with this trajectory.  A method for analyzing the past of a quantum particle according to the weak trace it leaves is proposed. Such a trace can be observed via measurements performed on an ensemble of identically pre-  and post-selected particles. Examples, in which this method contradicts the above common sense description of  the past of the particle are presented. It is  argued that it is possible to describe the past of a quantum particle, but the naive approach has to be replaced by both forward and backward evolving quantum states.
    \end{abstract}
\maketitle

\section{I. Introduction}

Contrary to classical physics, making a measurement and finding a quantum particle in a particular state does not tell us that this was its state in the past; it could have been in a superposition with some other states.  This explains why
 the fathers of quantum mechanics preached that we cannot talk about a quantum particle between measurements. Nevertheless, there is an extensive discussion of welcher weg (which path) detectors in the context of ``complementarity'' and other fundamental aspects of quantum mechanics \cite{Hel,Scu,Sto,Her,Durr,PRL1}, which all implicitly consider the past of a quantum particle. In this paper I argue  that we {\it can} describe the past of a quantum particle using  an objective criteria, but that this description contradicts the ``common sense'' approach widely used   in  analyses of which path interferometric experiments.

 The plan of the paper is as follows. In section II I analyze  Wheeler's delayed choice experiment using the ``common sense'' argument. In Section III I propose operational definition of the past of a quantum particle and show that it agrees with the common sense interpretation  in the case of Wheeler's experiment. Section IV describes another setup in which my proposed criterion for the past and the ``common sense'' argument contradict each other.  Section V presents the two-state vector formalism (TSVF) \cite{ABL,AV90} which provides the picture of the past of a quantum particle which is consistent with the criterion proposed in Section III. Section VI is devoted to welcher weg measurements involving additional degree of freedom. In section VII I analyze the results of the paper in frameworks of different interpretations of quantum mechanics. In Section VIII I summarize my conclusions.

\section{II. Delayed choice experiment}

   A natural approach to the past of a quantum particle was put forward by Wheeler \cite{Whe}. It asserts that while we cannot discuss the past of a particle until it is measured,  we can do so after the measurement. If the pre-selection led to a superposition of a few states and one of them was found in the post-selection measurement, then we should regard the particle as being in the post-selected state even before the post-selection.
Thus,  the past of a quantum particle comes into existence due to a measurement at a later time.

Recently an ``almost ideal'' experiment realizing  Wheeler's proposal for a delayed choice experiment has been performed \cite{Asp}. A single photon source was used and the choice of measurement was indeed delayed until the photon was inside the interferometer. The experiment demonstrated the bizarre feature of a quantum particle that a choice of measurement  performed now determines its behavior in the past. We can decide now, by removing or leaving the second beam splitter (BS) of the Mach-Zehnder interferometer (MZI), that the photon's past  is a single trajectory which is one of the arms of the interferometer, Fig. 1a or Fig. 1b, or that it is a superposition of both,  Fig. 1c.

\begin{figure}[h!tb]
  \includegraphics[width=5.5cm]{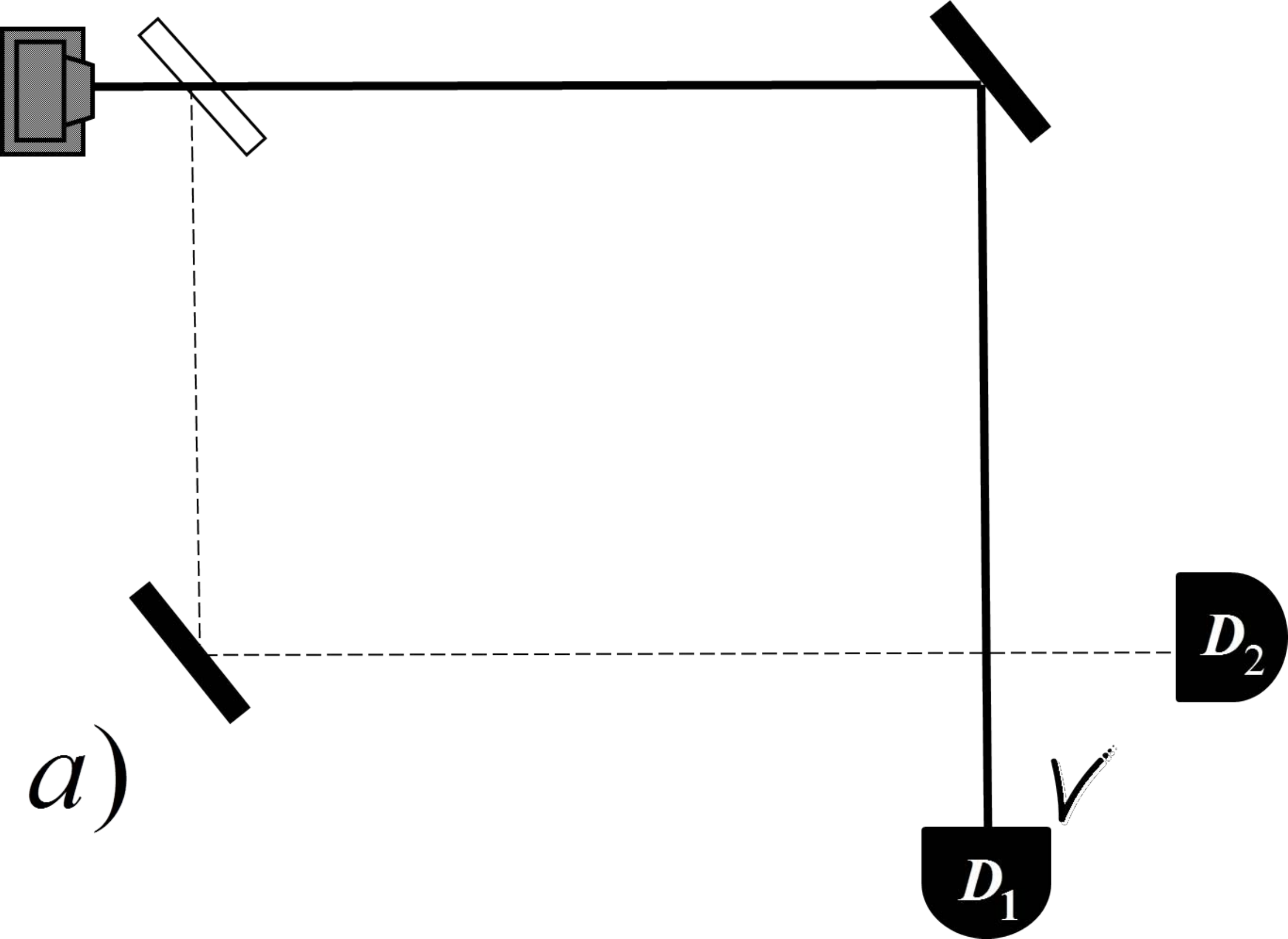}\\
  \includegraphics[width=5.5cm]{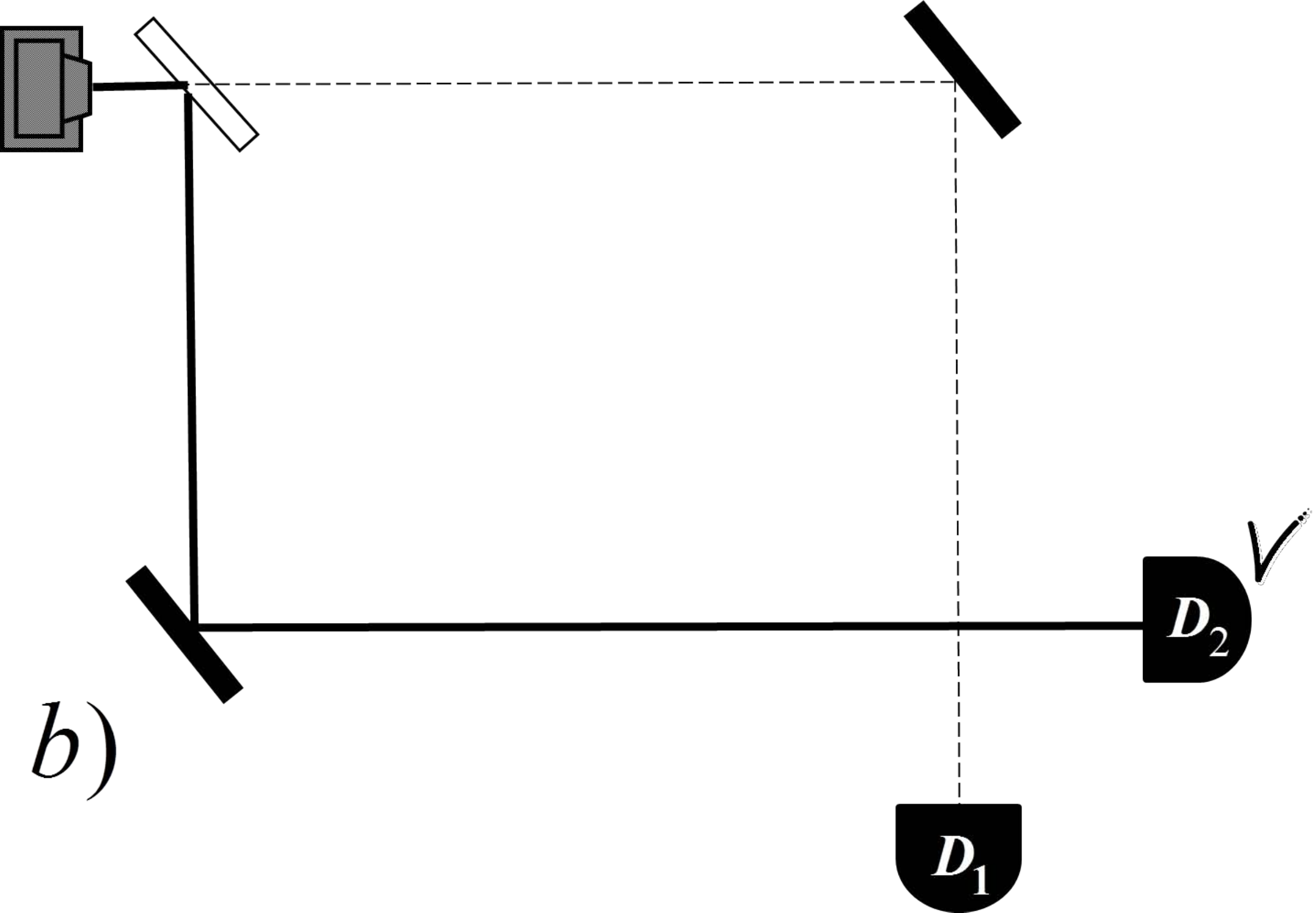}\\
  \includegraphics[width=5.5cm]{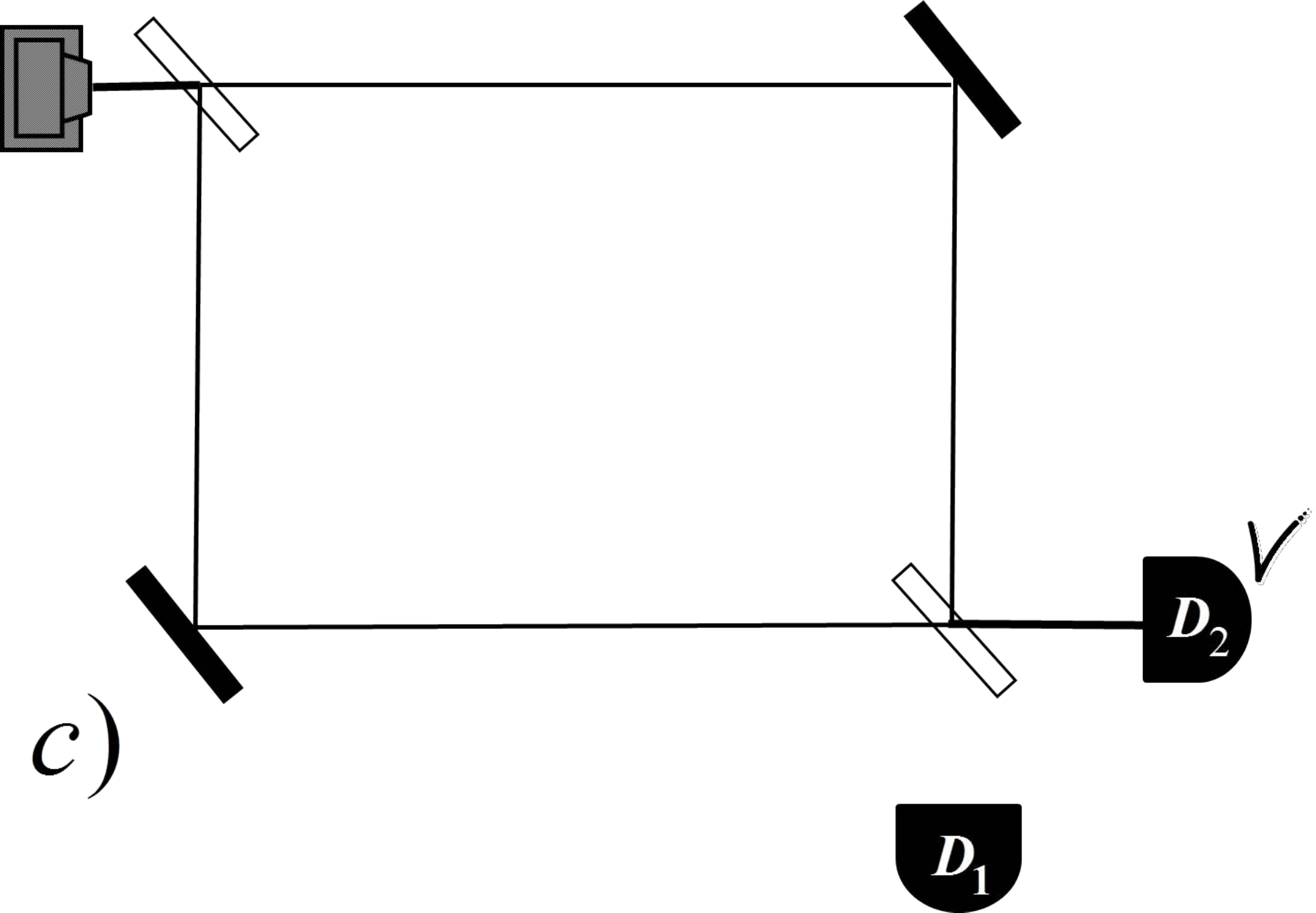}\\
  \caption{{\bf Wheeler's   delayed choice experiment.} Removing the second BS of the MZI causes the past of the photon to be a single trajectory, {\bf (a)} or {\bf (b)}, while leaving the BS forces the photon's past to be  a superposition of the two paths, {\bf (c)}. } \label{2}
\end{figure}

The experiment, however, demonstrated  this only indirectly. When the second  BS was missing, Fig. 1a. and Fig. 1b. it was considered as welcher weg measurement: if detector $D_1$  clicked  we concluded that the particle passed along the upper path because it could not reach the detector in another way and, similarly, if detector $D_2$  clicked  we  concluded that the particle passed along the lower path.  When the second BS was present,  we  concluded that the particle passed through both arms, since otherwise the interference (the dependence of the ratio of clicks of detectors $D_1$ and $D_2$  on  the phase of the interferometer) could not be explained.

\section{III. A criterion for the past of a quantum particle}

A direct manifestation of the past of a quantum particle would be an observation of the trace it leaves along its path. Arranging a strong measurement of the presence of a particle inside  the interferometer will not serve the purpose, because  that clearly changes the setup and the experiment becomes very different from the original proposal. (And surely we  cannot use  measurements such as the runs of the experiment with blocked arms of the interferometer \cite{Asp}.)

I propose  a criterion for the past of a quantum particle to be the weak trace it leaves. The particle interaction should be  weak enough such that the interference pattern is not disturbed significantly. An individual run does not provide enough information to reveal the past and we will need a large enough ensemble of identical experiments, the experiments in which particles were pre- and post-selected in the same states. We assume here that the setup, the pre-selection, and the post-selection provide complete description of the particle and, therefore, the weak trace observed on the ensemble  shows us the weak trace of each particle in the ensemble.

All particles have some nonvanishing interaction with the environment and they leave  some trace. In order to perform a quantitative analysis I will consider a particular model in which  weak von Neumann measurements \cite{VoN} are performed in both arms of the interferometer in Fig. 1.
The initial state of the pointers is a Gaussian with width  $\Delta $ centered at zero and the weak interaction leads (if the particle is there) to a  small shift
   \begin{equation}\label{shift}
\delta = \epsilon\Delta, ~~~~~~\epsilon \ll 1.
\end{equation}
  Performing such a measurement on a pre- and post-selected ensemble $N$ times will allow measuring  the shift with precision of  $\Delta \over \sqrt N $, so for $N > {1\over \epsilon^2}$ the presence of the particle will be revealed.

The outcomes of such measurements clearly support  the common sense picture. In the experiment with the second BS removed, in the ensemble with  $D_1$ detecting the particle, Fig. 1a, a shift $\delta$ will be observed at the upper arm of the interferometer and a shift zero at the lower arm. Of course, the reverse observations are obtained for the ensemble detecting the particle at $D_2$, Fig. 1b.

In the experiment with the second BS present and  tuned in  such a way that, due to the interference, the particles are detected only by $D_2$, Fig. 1c, the weak measurements in both arms of the interferometer  will show a shift of $ \delta \over 2$ manifesting an expected presence of the particles in both arms of the interferometer.

Note that the weak trace left by each particle has a particular correlation property. If, after the measurement interaction, instead of measuring the shift of the pointer position in one arm, we perform a projective measurement of its initial state,  then we will sometimes know with certainty that the particle passed through this arm. In those cases, a similar measurement performed in the other arm will not find the particle with certainty. This observation (which can be traced back to Einstein \cite{Ein}) can be interpreted in different ways depending on the adopted interpretation of quantum mechanics. Here we rely on the measurements on the ensemble for discussing the properties of each particle.  The modification of the reading procedure of the pointer of the measuring device transforms it to a (probabilistic) strong measurement and thus makes it inappropriate for observing the past of the particle inside the interferometer.

\section{IV. Nested Mach-Zehnder interferometer}

Let us now consider the past of a particle in the following modification of the MZI, Fig. 2. Instead of removing the second BS we add two   beam splitters and a mirror, ``nesting" another MZI inside the first one. (A similar proposal was used in an attempt to devise a setup for a counterfactual computation \cite{Ho,CFVA}.) The inner interferometer is tuned in such a way that there is a complete destructive interference in the lower output port, so that the wavepackets leaving the first BS and beginning to travel  in the two arms of the interferometer do not overlap thereafter.  The experiment apparently is a {\it welcher weg} measurement of the large interferometer: if the particle is detected by $D_3$, its past was the upper path of the large interferometer which includes both paths of the inner interferometer, Fig. 2a, while if detector $D_1$ or $D_2$ clicks, the past of the particle was the lower path of the large interferometer, Fig. 2b. In all these cases we made our claims relying on the fact that the particle could not reach the detectors via other paths.

\begin{figure}[h!tb]
  \includegraphics[width=6.1cm]{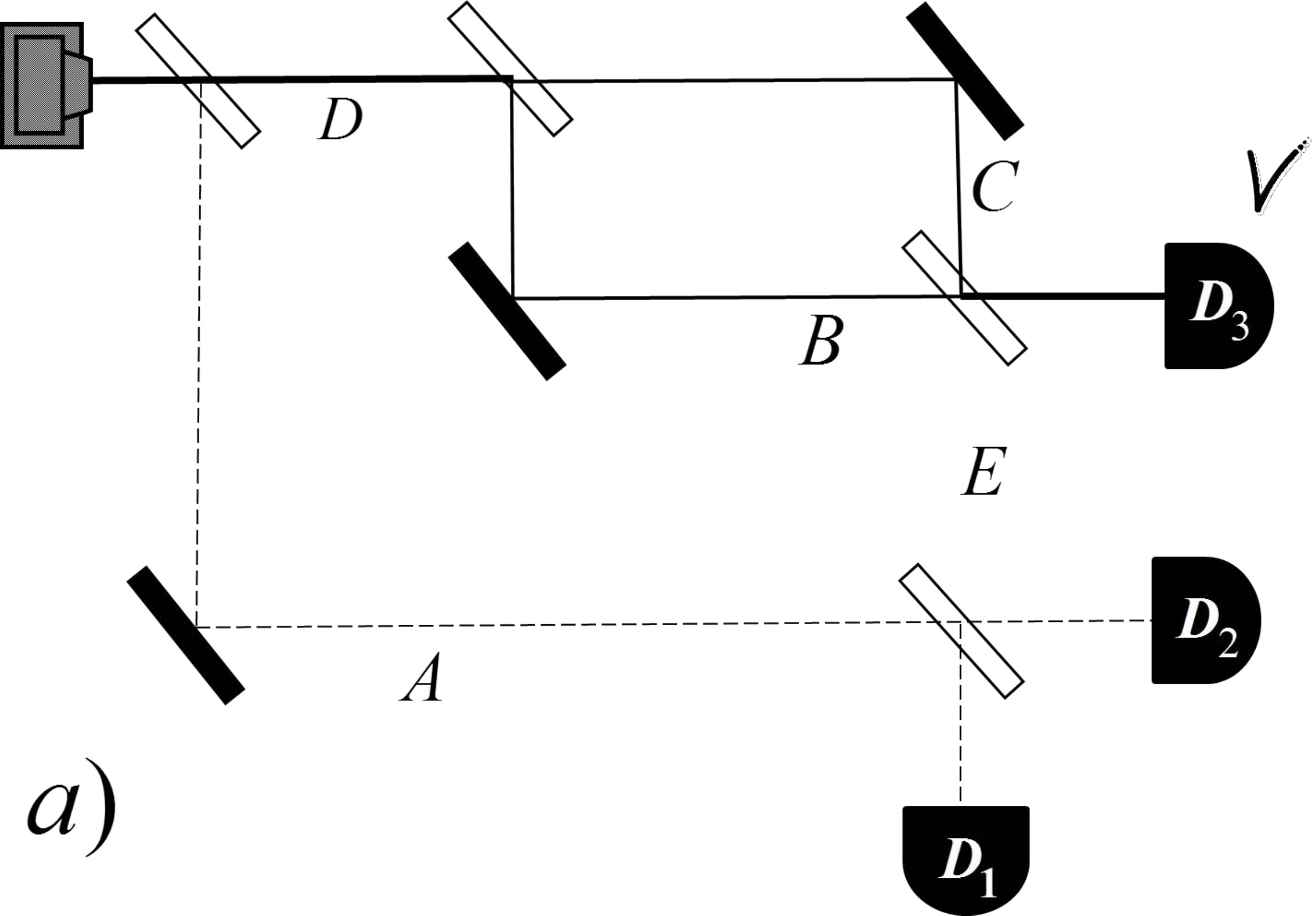}\\
   \includegraphics[width=6.1cm]{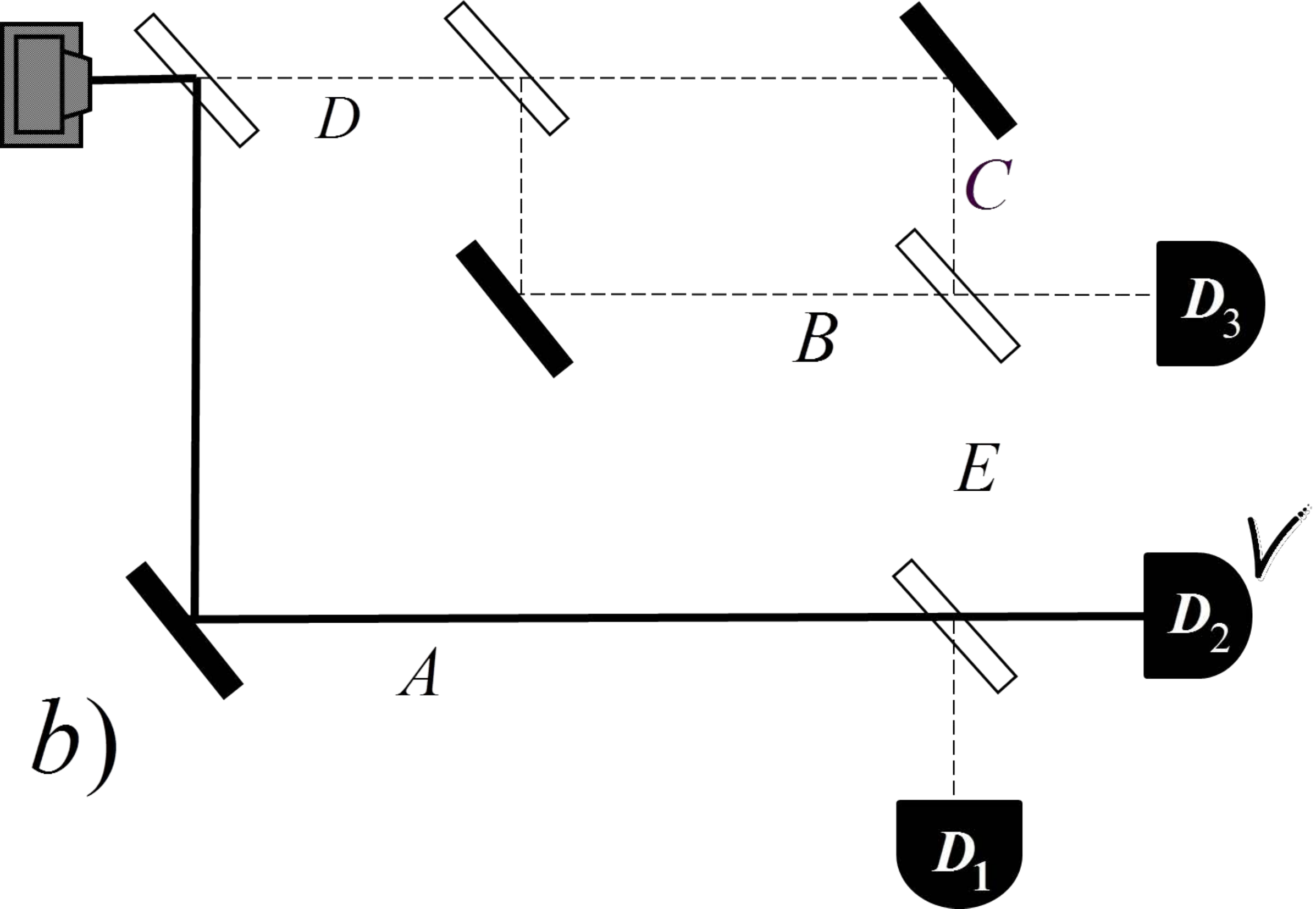}\\
   \caption{{\bf MZI nested inside another MZI.} The inner interferometer  is tuned in such a way that there is a destructive interference toward the lower output port. {\bf (a)} The ``common sense'' past of the  particle  if $D_3$ clicks. {\bf (b)} The ``common sense'' past of the  particle if $D_2$ clicks.} \label{4}
\end{figure}

The weak measurement tests for the path of the particle when detector $D_3$ clicks  shows that, indeed, it took the upper path of the large interferometer:
 \begin{equation}\label{shiftsD3}
\delta_D =1,~~~~~ \delta_B = \delta_C = {\delta \over 2}, ~~~~~~\delta_A =\delta_E =0.
\end{equation}
When detector $D_1$ or $D_2$ clicks, the weak measurement tests for the presence of the particle at points $A, D$ and $E$   also show the expected result. However,  when we weakly measure the presence of the particles inside the inner interferometer at points $B$ and $C$, a surprise happens. It seems that there should be no effect whatsoever since the particle could not have been there, but we see effects of the same order of magnitude, Fig. 3:
\begin{equation}\label{shiftsD12}
\delta_A =1,~~~
\delta_B = {\delta \over 2}, ~~~\delta_C = -{\delta \over 2}, ~~~~\delta_D =\delta_E =0.
\end{equation}

Weak measurements show that the photon leaves a trace in a path which it did not pass, or that our device is not a faithful  {\it welcher weg} measurement of the large interferometer in spite of the ``common sense'' argument: the photon detected in $D_1$ or $D_2$ could not pass through points $B$ and $C$. But if did pass through $B$ or $C$, why it did not leave a trace in $D$ and $E$? How this weak trace can be understood?

\begin{figure}[h!tb]
  \includegraphics[width=6.1cm]{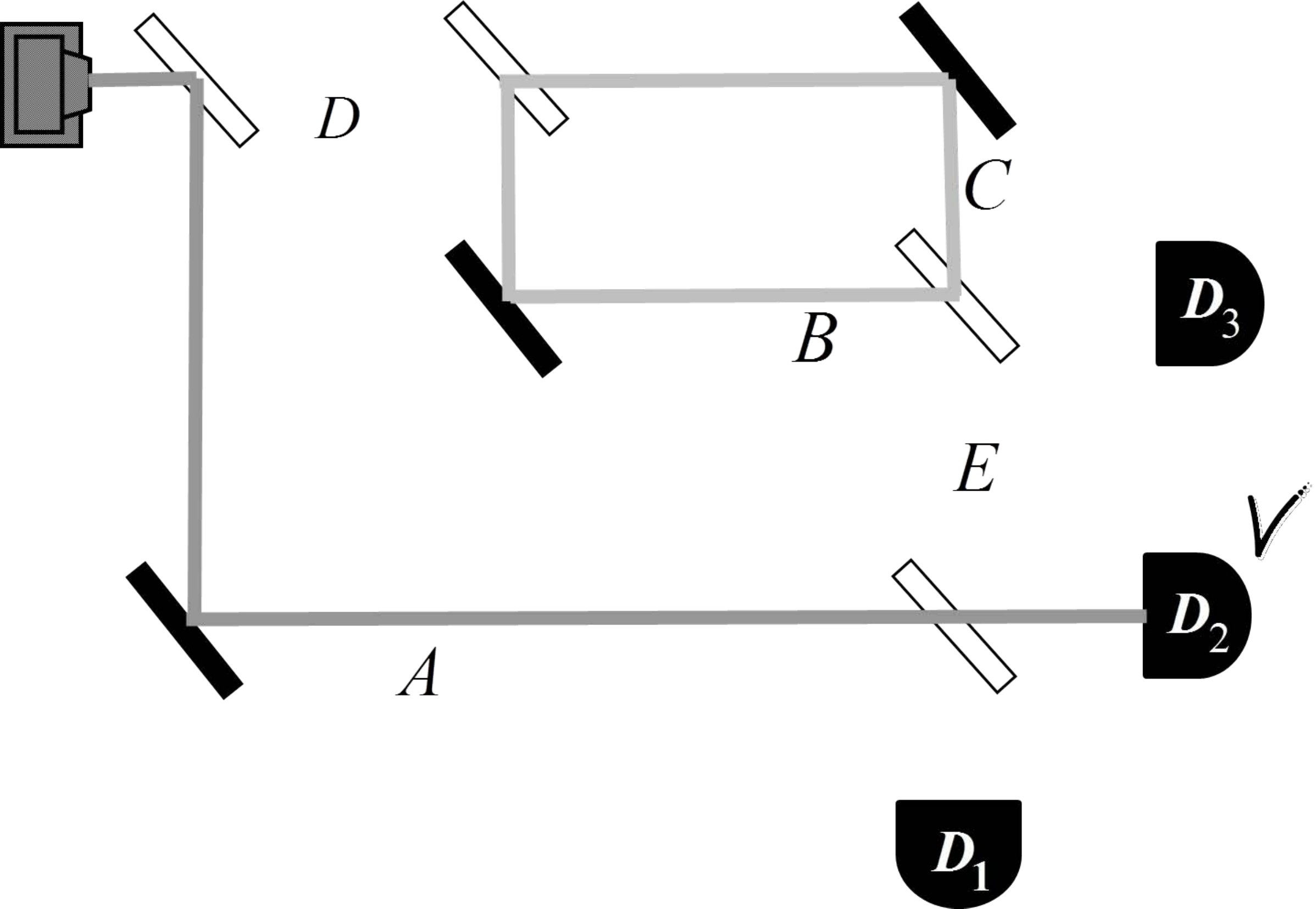}\\
      \caption{{\bf The past of the particle detected by $D_2$ according to the weak trace it leaves.} In addition to the continuous trace in the lower arm of the interferometer the particle leaves a similar strength trace  inside the inner interferometer which is a part of the upper arm of the large interferometer. } \label{7}
\end{figure}

\section{V. The two-state vector formalism}

The peculiar picture of the weak trace in the setup of the previous section is best viewed in the framework of the TSVF of quantum mechanics \cite{ABL,AV90}. According to this formalism,  a quantum system between two measurements is described by a two-state vector
\begin{equation}\label{tsv0}
  \langle{\Phi} \vert ~  \vert\Psi\rangle,
\end{equation}
which consists of the usual quantum state  evolving forward in time, $ |\Psi\rangle $, defined by the
results of a complete measurement at the earlier time, and by a quantum state evolving backward in time $ \langle \Phi |  $, defined by the results of a complete measurement at a later time. Any weak coupling to a variable $O$ of a pre- and post-selected quantum system results in an effective coupling to a {\it weak value} of this variable:
\begin{equation}\label{wv}
O_w \equiv { \langle{\Phi} \vert O \vert\Psi\rangle \over
\langle{\Phi}\vert{\Psi}\rangle }  .
\end{equation}
We expect that, due to the locality of all interactions, the trace the particle leaves is proportional to the weak value of the particle's projection operator onto a  particular location. Thus, the shift of the pointer of the weak measurement device is proportional to the weak value of the projection operator of the particle at the location of the device. It vanishes for locations where the forward or backward evolving state vanishes. Figs. 4a,b show how this picture explains Wheeler's conclusions for his experiment, while Fig. 4c shows how it explains the peculiar results in the modified setup.
 In the places where both waves are present, the weak value quantifies the shift of the pointer of the weak measuring device. The two-state vector of the photon inside the interferometer at a particular moment  is
\begin{equation}\label{tsv}
 \nonumber
\langle{\Phi} \vert ~  \vert\Psi\rangle =
{1\over \ 2}\left(\sqrt 2 \langle A \vert +  \langle B \vert  -   \langle C \vert\right )  ~~ {1\over \ 2}\left(\sqrt 2 \vert A \rangle +   \vert B \rangle +  \vert C \rangle \right ) ,
\end{equation}
where we used natural notation: $\vert A \rangle$ is a localized wave packet at point $A$, etc. Then, (\ref{wv}) yields for the weak values of the projection operators at points  $A, ~B$ and  $C$:
\begin{equation}\label{wvs}
 ({\rm \bf P}_A)_w =1,~~~~~({\rm \bf P}_B)_w ={1\over 2},~~~~~({\rm \bf P}_C)_w =-{1\over 2} .
\end{equation}
This explains the shift of weak measuring devices (\ref{shiftsD12}) placed at these points.

 A variation   of such a measurement  in which the measuring device is a separate degree of freedom of the particle itself (its lateral position) has been performed \cite{resh}.  In this experiment a thin crystal plate at some angle was placed at various points and the lateral shifts were measured in the final detection. Measurements have been performed separately at  points  $A, ~B$ and  $C$ and shifts as in (\ref{shiftsD12}) were obtained. However, the vanishing shifts at points  $D$ and  $E$ have not been tested.

\begin{figure}[h!tb]
  \includegraphics[width=6.1cm]{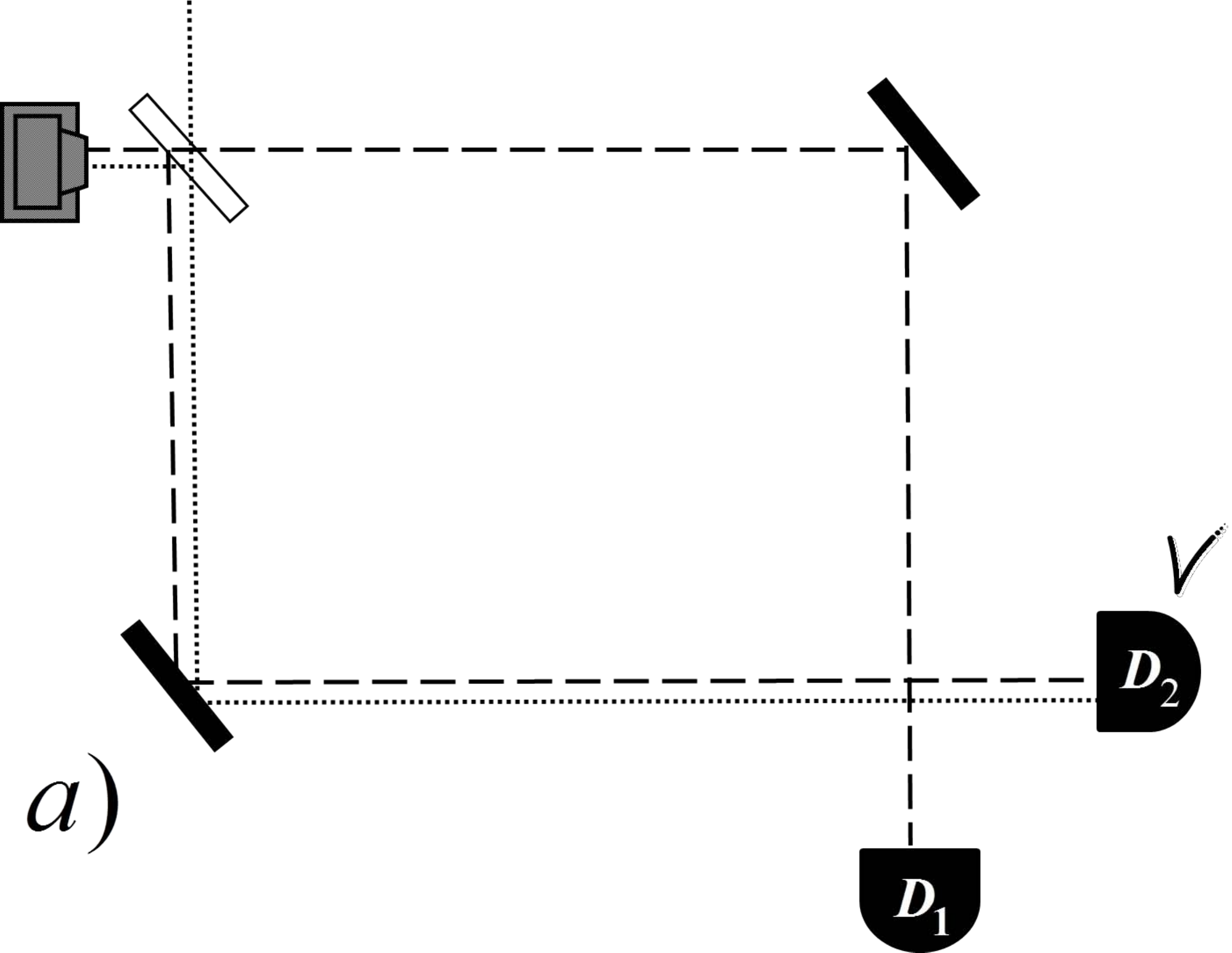}\\
   \includegraphics[width=6.1cm]{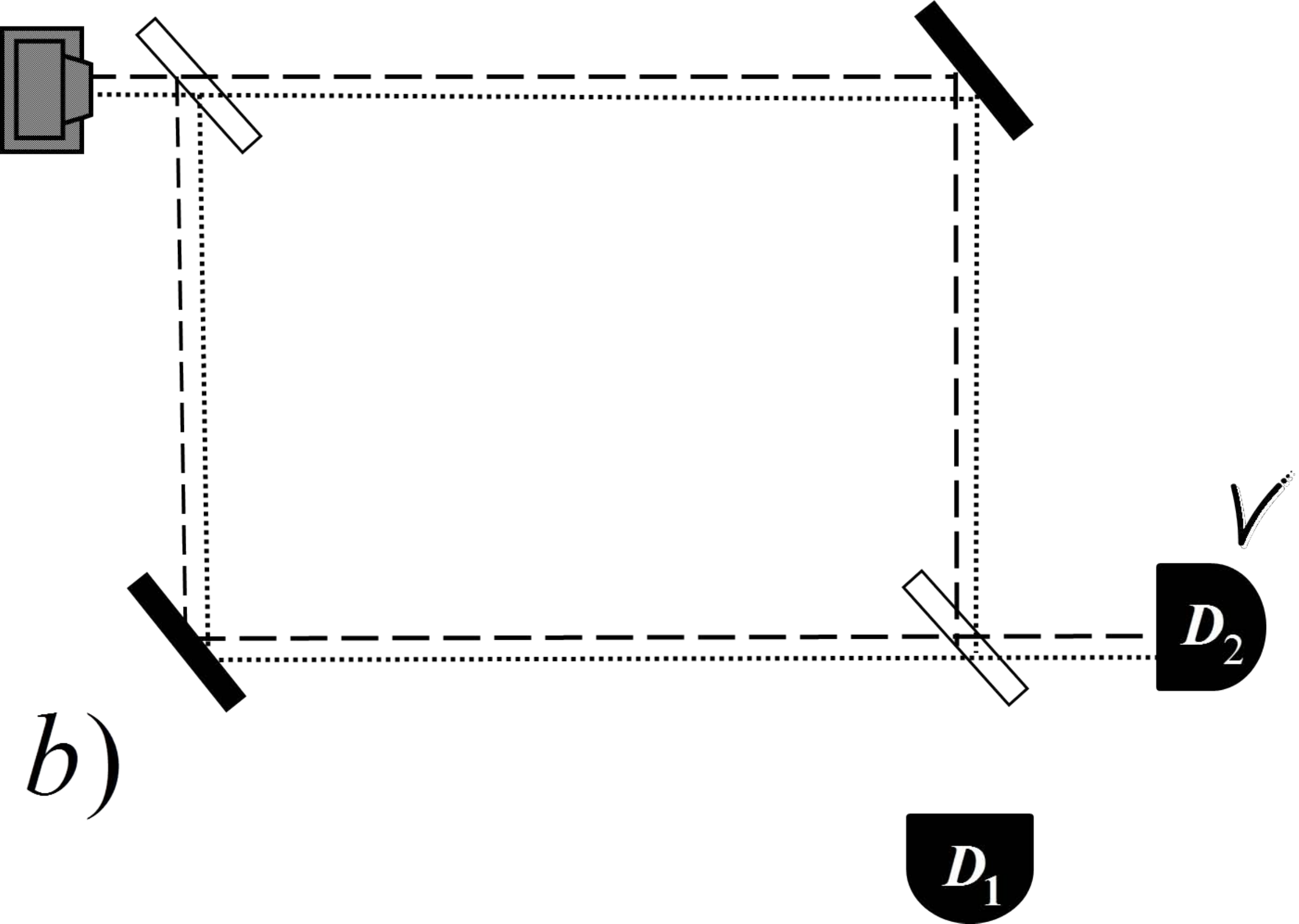}\\
    \includegraphics[width=6.1cm]{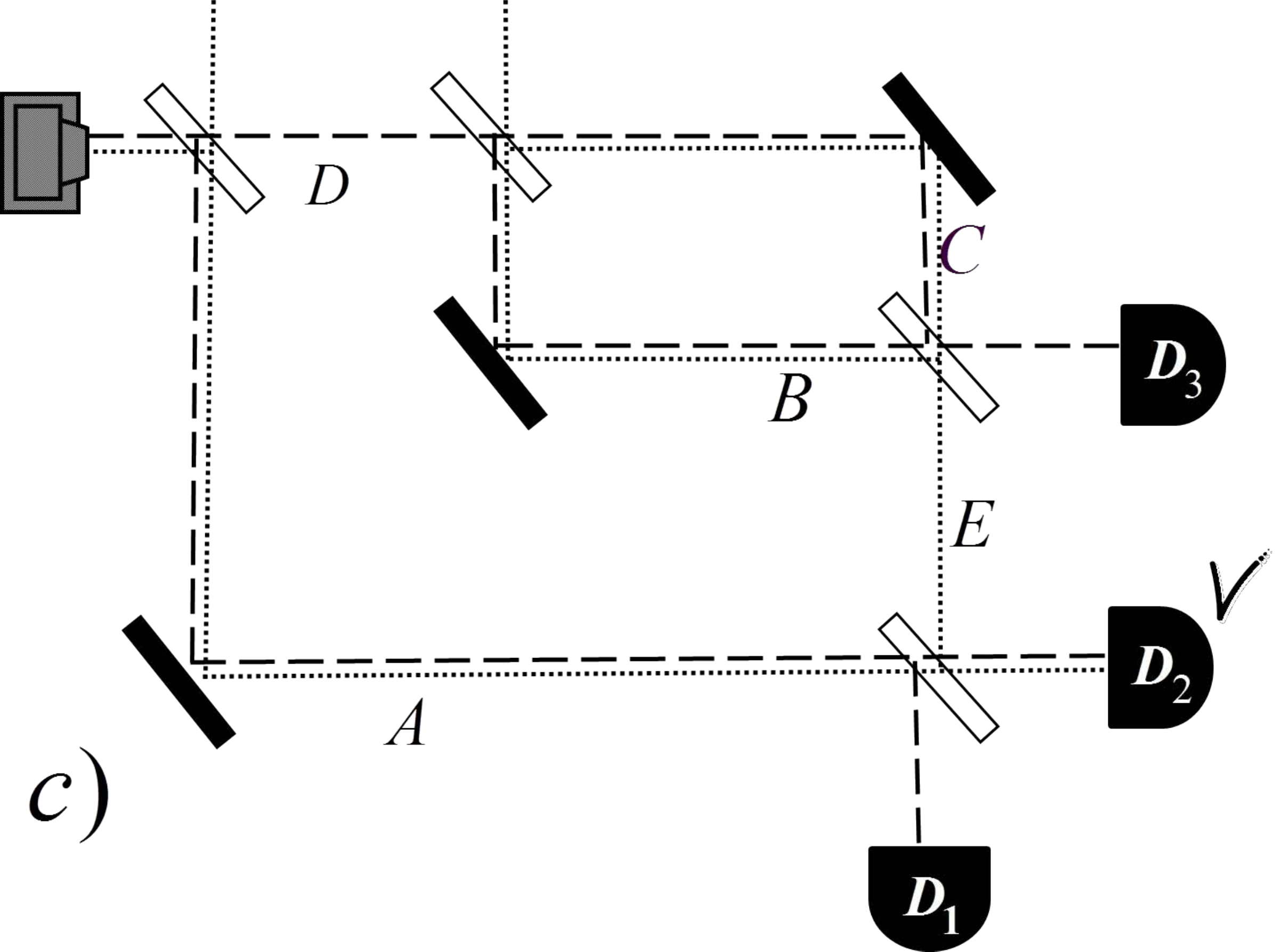}\\
   \caption{{\bf The TSVF description.} The particle leaves a trace only where both forward (dashed line) and backward (dotted line) evolving states are present. {\bf a)} Wheeler's experiment without the second BS, compare with Fig. 1b; {\bf b)} Wheeler's experiment with the second beam splitter in place, compare with Fig. 1c; {\bf c)} Modified experiment, compare with Fig. 3.} \label{8}
\end{figure}

The fact that  the weak coupling inside the inner interferometer when a particle  arrives at detector $D_1$ or $D_2$ leads to an observable shift, is very peculiar. In the standard formalism of quantum mechanics it can be explained as  a counterintuitive interference effect.
 The weak coupling to the measuring device slightly spoils the destructive interference and a tiny wave goes toward the second BS of the large interferometer. The amount of the wave which is leaked out is very  small. The ratio of this flux and the flux at the lower arm of the large interferometer is
 \begin{equation}\label{ratio}
 {{{1\over 2}\int_{-\infty}^{\infty} (e^{-{{x^2}\over {2\Delta^2}}}- e^{-{{(x-\delta)^2}\over {2\Delta^2}})^2}dx}
 \over {\int_{-\infty}^{\infty} e^{-{{x^2}\over \Delta^2}}dx}} = {{(1-e^{-{\delta^2\over{4\Delta^2}}})}\over2} \simeq {{\epsilon^2}\over 8}.
\end{equation}
 Nevertheless, due to the amplification effect of post-selection, the shift of the measuring device is of the first order in $\epsilon$. It is comparable with the shift caused by a particle which is fully present near the measuring device.

\section{VI. {\it Welcher weg}   measurement with polarization}

Consider a variation of a {\it welcher weg}   measurement in a MZI (it has been extensively analyzed in connection with ``quantum eraser" experiments \cite{Her}).  We start with horizontally  polarized photon $|H\rangle$ and we ``mark" the photon in the upper arm $B$ by rotating the polarization to vertically polarized state $|V\rangle$, see Fig. 5. Assume that at the end the photon was found in polarization state $|H\rangle$. Then, since in the arm $B$ there were no photon in state $|H\rangle$, the ``common sense'' approach tells us that the photon passed through lower arm $A$.

\begin{figure}[h!tb]
  \includegraphics[width=6.9cm]{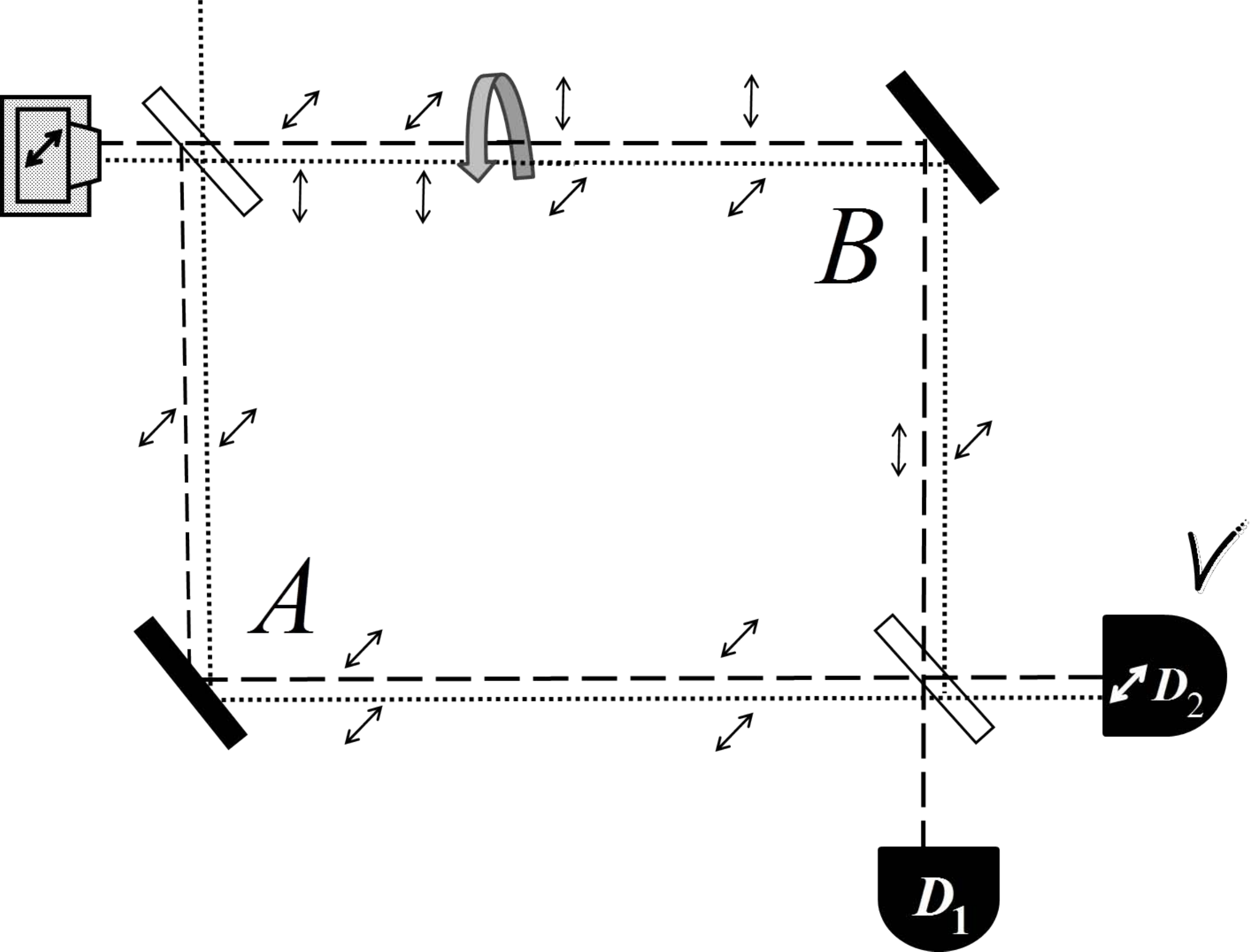}\\
    \caption{{\bf Welcher weg measurement.} The photon polarization in the MZI with polarization insensitive BSs is rotated to an orthogonal state in the   arm $B$. The dashed line shows forward evolving quantum state of the photon emitted  in  $|H\rangle$ state, while dotted line shows backward evolving state of the photon detected in  $|H\rangle$ state.} \label{9}
\end{figure}
 The two state vector of the quantum particle inside the interferometer is:
\begin{equation}\label{state}
 \langle{\Phi} \vert ~  \vert\Psi\rangle =
{1\over \sqrt 2} \left ( \langle A \vert +  \langle B \vert \right ) \langle H \vert   ~~ {1\over \sqrt 2} \left(\vert A \rangle  \vert H \rangle+ \vert B \rangle  \vert V \rangle \right).
\end{equation}
Thus, the weak values (\ref{wv}) for projection operators in the arms of the interferometer  support this conclusion:
\begin{equation}\label{PO1}
({\rm \bf P}_B )_w = 0,~~~({\rm \bf P}_A )_w = 1.
\end{equation}
 Nevertheless,  the weak trace might show otherwise. Indeed, the  basic constraint  on interactions in nature is that they should be local. It is possible to have local coupling to, say, clockwise polarized photon. Such coupling will lead to identical traces in the lower and upper arm of the interferometer:
  \begin{equation}\label{PO2}
({\rm \bf P}_B {\rm \bf P}_\circlearrowright )_w = ({\rm \bf P}_A {\rm \bf P}_\circlearrowright)_w ={1\over 2}.
\end{equation}
The criterion of the weak trace tells us that the particle has been in both arms of the interferometer. Again, the TSVF explanation works: the photon  was  where both forward and backward evolving wave functions do not vanish, Fig. 5.

In another very similar modification of the welcher weg measurement the picture is different. Consider theoretically possible, but hardly feasible with today's technology experiment in which the photon in arm $B$, instead of rotating its own polarization, changes polarization (or spin) of another particle to an orthogonal state. Consider a post-selection according to which the photon is detected at $D_2$ and also at that time the polarization of the other particle is observed to be in its original state.  Mathematically, this situation is also described by the two-state vector of the form (\ref{state}), only now it is the two-state vector describing the composite system of the photon and the external particle (which is spatially separated from the photon) where $\langle H \vert$,  $\vert H \rangle$ and $\vert V \rangle$ signify polarization states of the other particle.  In this experiment the weak trace of the photon in the arm $B$ is zero.  Due to orthogonality of the polarization state of the external particle the weak values of all local photon observables in the arm $B$  vanish.

When two particles are pre- and post-selected in entangled states, each particle itself is described by   a {\it generalized two-state vector} \cite{AV91}. In our situation  only the pre-selected state is entangled, so the photon is described just by a simple  two-state vector:
 \begin{equation}\label{state}
 \langle{\Phi} \vert ~  \vert\Psi\rangle =
{1\over \sqrt 2} \left ( \langle A \vert +  \langle B \vert \right )    ~~ \vert A \rangle.
\end{equation}
 It explains why we have weak trace only in arm $A$.

 In all our examples the TSVF provides a very simple and natural explanation of the weak trace of the quantum particle. We can see the weak trace   only where we have nonvanishing forward and backward evolving quantum states.

\section{VII. Analysis  in the frameworks of different interpretations}

Let us discuss how different interpretations of quantum mechanics treat the past of a quantum particle and explain the weak trace the particle leaves in the examples above.

The simplest approach is, following Bohr, to attribute reality only to measurement outcomes, thus disallowing discussion of the past of the particle even after it has been measured. Indeed, Bohr forbids looking  for a description of the particle between the pre- and post-selection. In this approach the weak trace can be explained only if the quantum analysis of everything involved (the particle, environment, measuring devices, etc.) will be performed.

The textbook approach  postulating   collapse of the quantum state at measurement (which might be attributed to von Neumann \cite{VoN}), does provide a picture of the past. However, this picture is independent of the outcome of the post-selection measurement. Between the pre- and post-selection, the quantum particle in the interferometer  is described by the pre-selection state only. In our examples the quantum wave is present in all arms of interferometers. This  picture in most cases does not represent the weak trace the particle leaves (the experiment described in Fig. 1c. is an exception). Thus, the von Neumann description of the particle alone is not sufficient to explain the weak trace.

In the de Broglie-Bohm interpretation (DBB) \cite{DB,Bohm52} the ontology consists of noncollapsing quantum wave of a quantum particle and its position. The proponents  of the DBB usually consider the  latter  as the {\it primary} ontology. In many situations, the   pre- and post-selection of the state of the particle specifies the DBB past exactly, i.e. it tells us what was the Bohmian trajectory of the particle \cite{BoDeHi}.

In my central example described in  Fig. 2b. in which the weak trace contradicts the ``common sense'' picture, the DBB path is the same as the common sense tells us. It has to be the lower path because the DBB particle position must ride on a nonvanishing wave packet and there is not wave packet passing through the upper arm. However, the DBB picture does not always agrees with the common sense. Indeed in the case presented  in Fig. 1a. (as well as in Fig. 1b.), the DBB path is just the opposite of the ``common sense'' path. The Bohmian particle ``rides'' the wavepacket moving in the lower arm and it switches to the wave packet of the upper arm when  the wave packets overlap \cite{Bell}.

The DBB trajectories do not explain the weak trace neither in case of Fig. 1a. nor in case of Fig. 2b. In fact, these are not the first examples of discrepancy between DBB picture and weak measurements of position  \cite{BohmWeak}. This discrepancy, however, is not very surprising because in similar setups with particular kinds of position  measurements the particles can leave even a strong trace far away from the DBB trajectories \cite{sur,gili}.

As in the interpretations mentioned above, the lack of an explanation of the weak and even of the strong trace left by a quantum particle via particle's description in the DBB interpretation does not make the theory inconsistent. The explanation can be made, but it requires consideration of other systems too.

The description of a quantum particle in a particular world in the many-worlds interpretation (MWI) \cite{Everett} between the pre-selection and the post-selection seems to be  identical to that of von Neumann: collapse at measurement is just replaced by splitting at measurement, and the particle in an interferometer is described by the quantum state specified by pre-selection only. This description does not provide  an explanation of  the weak trace of the quantum particle.

According to my approach to the MWI \cite{SEP} ``a world'' of the MWI is a sensible story with causal connections. Hence, this apparent inability to explain the above phenomena seems to be an inconsistency. This led me to propose a modification of the MWI \cite{E50} in which weakly interacting particles at the intermediate time between two measurements (such as a photon in an interferometer) are described by the two-state vector which includes the information of the post-selection measurement and explains the weak trace. (There is no need to describe strongly interacting systems by backward evolving state because it is identical to the forward evolving state.)

With this modification, I find the  TSVF to be very natural in the framework of the MWI. A two-state vector of a particle with its corresponding weak trace is a well defined concept in a particular world. In my world with the post-selection I have seen, I should also expect to have the effects of the corresponding weak trace. The surprising picture of the weak trace described in Fig. 3., where there is a trace inside the inner interferometer, but there is no trace leading towards it, is the unique feature of the world with a click in $D_2$. God, or super technology which  observes the effects of all outcomes in parallel, will see a continuous trace in all arms of the interferometer (except $E$), the trace predicted by the forward evolving quantum state.

It will be interesting to see if the weak trace in the examples presented above have natural explanation in other interpretations, such as physical collapse theories \cite{GRW,Pearle}, consistent histories approach \cite{Grif}, Nelson-Guerra stochastic mechanics \cite{Nel,Gue}, but this goes beyond the scope of this paper.

\section{VIII. Summary and Conclusions}

I have   shown that the ``common sense'' approach to the past of a quantum particle does not always correspond to the weak trace it leaves on the environment, specifically,  on a specially designed weak measurement device.

One may learn  different lessons from this observation depending on his/her favorite interpretation of quantum mechanics. One can, following Bohr,  refuse to talk about the past and thus avoid inconsistencies, but in the process loose a useful insight. Alternatively, one can following Bohm  construct a consistent deterministic picture of the world and admire a sophisticated nonlocal mechanism responsible for the trace observed far away from particles trajectories.
My preference is the many-worlds interpretation and my  lesson is  the necessity of  a slight modification of  Everett's concept of a world \cite{E50}. Description of  the reality of a quantum particle in a particular world requires both forward and backward evolving quantum states. The two-state vector provides ``weak-measurement reality'' \cite{WMER} for quantum particles between measurements based on the weak value (\ref{wv}) in a world with a particular pre- and post-selection.  This ``reality" is not just a theoretical construction since it can be demonstrated with current technology.

One can learn useful lessons even without attempting to apply physics for description of reality. First, the peculiar effect presented here teaches us to use ``common sense'' with care: the approach, according to which we decide about the past of a quantum particle based on the fact that ``it could not come the other way", has to be abandoned.   Second, the description of the quantum particle itself which correctly describes the weak trace it leaves is the two-state vector formalism \cite{ABL,AV90}.

\section*{Acknowledgements}

This work has been supported in part by  grant number
32/08 of the Binational Science Foundation and the Israel Science Foundation  Grant No. 1125/10.


\begin{thebibliography}{99}
\bibitem{Hel}
T. Hellmuth, H. Walther, A. Zajonc, and W. Schleich,
Phys. Rev. A {\bf 35}, 2532 (1987).
 \bibitem{Scu}
 M. Scully, B. Englert, and H. Walter,
Nature (Londoon) {\bf 351},    111  (1991).

  \bibitem{Sto}
  P. Storey, S. Tan, M. Collett, and D. Walls,
  Nature {\bf 367},    626  (1994).

\bibitem{Her} T.J. Herzog, P. Kwiat, H Weinfurter, and A. Zeilinger,
Phys.  Rev. Lett.  {\bf 75}, 3034 (1995).

 \bibitem{Durr}
 S. Durr, T. Nonn, and G. Rempe,
Nature (Londoon) {\bf 395},    33  (1998).

\bibitem{PRL1}
V. Jacques,  {\it et al.},  Phys. Rev. Lett.  {\bf 100},   220402 (2008).

\bibitem{ABL}
 Y. Aharonov, P. G. Bergmann,  and J. L.   Lebowitz,    Phys.  Rev. B {\bf 134}, 1410 (1964).

\bibitem{AV90}
Y. Aharonov  and L. Vaidman,
   Phys. Rev. A \textbf{41}, 11 (1990).


\bibitem{Whe}
J. A. Wheeler, ``The `past' and the `delayed-choice double-slit experiment'," in {\it Mathematical Foundations of Quantum Theory}, A.R. Marlow, ed., pp. 9–48,  (Academic Press 1978).



\bibitem{Asp}
V. Jacques, {\it et al}.,
 Science {\bf 315}, 966 (2007).

\bibitem{VoN}
J. von Neumann,    {\em Mathematical Foundations of Quantum Theory},
  (Princeton: Princeton University Press, 1955).

\bibitem{Ein}
A. Einstein, R.C. Tolman,  and B. Podolsky,  Phys. Rev. {\bf 37}, 780 (1931).

 \bibitem{Ho}
O. Hosten {\it et al.},
Nature (Londoon) {\bf 439}, 949 (2006).

\bibitem{CFVA}
L. Vaidman,
Phys.  Rev. Lett.  {\bf 98}, 160403 (2007).


\bibitem{resh}
K.J. Resch, J.S. Lundeen, A.M. Steinberg,
Phys. Lett. A {\bf 324}, 125 (2004).


\bibitem{AV91}
Y. Aharonov  and L. Vaidman,   J.  Phys. A {\bf   24}, 2315 (1991).


\bibitem{DB}
L.~de~Broglie,
\newblock {Une tentative d'interpretation causale et non lin{\'e}aire de
  la m{\'e}canique ondulatoire (la th{\'e}orie de la double solution).}
\newblock (Gauthier-Villars, Paris, 1956);
\newblock English translation: {\em Non-linear Wave Mechanics: A Causal
  Interpretation} (Elsevier, Amsterdam, 1960).

\bibitem{Bohm52}
D. Bohm, Phys. Rev.  {\bf 85}, 166 (1952).

\bibitem{BoDeHi}
D. Bohm, C. Dewdney, and B.  Hiley, Nature (Londoon) {\bf 315}, 294 (1985).


 \bibitem{Bell}
J. S. Bell,
Int. J. Quan. Chem. {\bf 14}, 155 (1980).

\bibitem{BohmWeak}
Y. Aharonov and L. Vaidman,
in {\it Bohmian Mechanics and Quantum Theory: An Appraisal}, J.T. Cushing, A. Fine and S. Goldstein (eds.), pp. 141-154 (Kluwer, Dordrecht, 1996).

\bibitem{sur}
B.G. Englert, M.O.~Scully, G.~S\"{u}ssmann, and H.~Walther,
 { Z. Naturforsch. A} {\bf 47}, 1175 (1992).

\bibitem{gili}
G. Naaman-Marom, N. Erez, and  L. Vaidman,  Ann. Phys. {\bf 327}, 2522 (2012).


\bibitem{Everett}H. Everett III,
 Rev. Mod. Phys. {\bf 29}, 454-462  (1957).

\bibitem{SEP} L.~Vaidman,  Many-Worlds Interpretation of Quantum
Mechanics, {\it Stan. Enc. Phil.},  E. N. Zalta (ed.) (2002),
http://plato.stanford.edu/entries/qm-manyworlds/.

\bibitem{E50}
L. Vaidman,
Time Symmetry and the Many-Worlds Interpretation,
in {\it Many Worlds? Everett, Quantum Theory, and Reality},
S. Saunders, J. Barrett, A. Kent, and D. Wallace eds., (Oxford University Press 2010).

\bibitem{GRW} G.C. Ghirardi, A. Rimini,  and T. Weber,
Phys. Rev. D {\bf  34}, 470 (1986).

\bibitem{Pearle}P. Pearle,
Phys. Rev. A {\bf  39}, 2277 (1989).


\bibitem{Grif}
R.B.~Griffiths, J. Stat. Phys. {\bf 36}, 219 (1984).

\bibitem{Nel}
E. Nelson,
Phys. Rev. {\bf 150}, 1079 (1966).

\bibitem{Gue}
F. Guerra, Phys. Rep. {\bf 77}, 263 (1981).


\bibitem{WMER}
 L. Vaidman,
 Found. Phys. {\bf 26}, 895 (1996).



\end{thebibliography}
\end{document}